\documentstyle[12pt]{article}
\begin{document}
\begin{titlepage}
\begin{center}
{\bf On the two-magnon bound states for the quantum Heisenberg chain
with variable range exchange}\\
\vspace*{1cm}
{\sc J. Dittrich}\\
{\it Nuclear Physics Institute, Academy of Sciences of the Czech Republic, 
CZ-250 68 \v Re\v z, Czech Republic
\footnote{Also member of the Doppler Institute of Mathematical Physics,
Faculty of Nuclear Sciences and Physical Engineering, Czech Technical 
University, Prague.}
}\\
and\\
{\sc V.I. Inozemtsev}\\
{\it BLTP, JINR, 141980 Dubna, Russia}\\
\vspace*{1.5cm}
Abstract\\
\end{center}
The spectrum of finite-difference two-magnon operator is investigated
for quantum $S$=1/2 chain with variable range exchange of the form
$h(j-k)\propto \sinh^{-2}a(j-k)$. It is found that usual bound state
appears for some values of the total pseudomomentum of two magnons as 
 for the Heisenberg chain with nearest-neighbor spin interaction.
Besides this state, a new type of bound state with oscillating wave function
appears at larger values of the total pseudomomentum.
\end{titlepage}

The spectral theory of finite-difference operators associated with multimagnon
states of one-dimensional ferromagnetic Heisenberg chains is far from being
complete. From previous works, it is known that magnons not only scatter on
each other, but can also create bound complexes with lower energy. A canonical
example of this phenomenon was described first by Bethe [1] for the famous
case of nearest-neighbor exchange. It was found that in this case just one 
bound state appears for all values of the total pseudomomentum $P$ in the
interval $0<P\leq \pi$. Much later, it was shown that for
next-nearest-neighbor interaction the bound state is always present within 
this interval but can be destroyed at $P=\pi$ by second-neighbor
interaction [2-4]. The situation for the third neighbor or more general 
nonlocal spin interaction has not been investigated till now.

In this letter, we shall investigate the two-magnon problem for the integrable
ferromagnetic Heisenberg chain with variable range exchange defined by the
Hamiltonian
$$H=-{J\over 2}\sum_{j\neq k;j,k=-\infty}^{\infty}h(j-k)
{{\vec \sigma_{j}\vec \sigma_{k}-1}\over 2},\eqno(1)$$
where $\{\vec \sigma_{j}\}$ are Pauli matrices attached to the site $j$,
 $J>0$ and
$$h(j-k)={1\over{\sinh^{2}a(j-k)}}.\eqno(2)$$
The range of variation of the parameter $a$ is $0<a<\infty$. It will be shown
that for all values of $a$ in this interval there are two types of two-magnon
bound states, the first is similar to the bound state for the Heisenberg chain
with nearest-neighbor interaction which might be considered as a limit of (1-2)
as $a\to\infty$ (with $J\propto \sinh^2 a$), and the second corresponds to the asymptotically decayed
two-magnon wave functions with superimposed oscillations.

We shall consider two-magnon excitations over the ground state of the model in
the form
$$\vert\psi>=\sum_{n_{1}\neq n_{2}}\psi(n_{1},n_{2})S_{n_{1}}^{-}S_{n_{2}}
^{-}\vert 0>,\qquad S_{n}^{-}={1\over2}(\sigma_{n}^{x}-i\sigma_{n}^{y}),$$
where $\vert 0>$ is the ferromagnetic ground state with all spins aligned up
and $\psi(n_{1},n_{2})=\psi(n_{2},n_{1})$.
Acting with (1) on the state $\vert\psi>$, the equation $H\vert\psi>
=\varepsilon\vert\psi>$ results in
$$H^{(2)}\psi(n_{1},n_{2})=\varepsilon\psi(n_{1},n_{2}),\eqno(3)$$
where the action of the finite-difference operator $H^{(2)}$ is given by
$$H^{(2)}\psi(n_{1},n_{2})=-[
\sum_{k\neq n_{1},n_{2}}(h(n_{1}-k)\psi(k,n_{2})+h(n_{2}-k)\psi(n_{1},k))
$$
$$\left.
+(2h(n_{1}-n_{2})-2\varepsilon_{0})\psi(n_{1},n_{2})
\right],\eqno(4)$$
where $\varepsilon_{0}=\sum_{k\neq 0}h(k)$. A two-parametric solution
to the eigenequation (3) with the exchange strength (2) reads as follows [5],
$$\psi(n_{1},n_{2})=e^{i(p_{1}n_{1}+p_{2}n_{2})}(\coth\gamma(p_{1},p_{2})
+\coth a(n_{1}-n_{2}))
$$
$$                 +e^{i(p_{2}n_{1}+p_{1}n_{2})}(\coth\gamma(p_{1},p_{2})
-\coth a(n_{1}-n_{2})),\eqno(5)$$
where the phase factor $\gamma(p_{1},p_{2})$ is expressed via the magnon 
pseudomomenta $p_{1,2}$,
$$\coth \gamma(p_{1},p_{2})={{f(p_{1})-f(p_{2})}\over {2a}}.\eqno(6)$$
The formula for two-magnon energy reads
$$\varepsilon(p_{1},p_{2})=
\varepsilon(p_{1})+
\varepsilon(p_{2})=J\sum_{n\neq 0}{{2-\cos(np_{1})-\cos(np_{2})}\over
{\sinh^2(an)}},
\eqno(7)$$
where the sum can be expressed via the Weierstrass functions.
The structure of the function $f(p)$ will be crucial for our treatment.
It is given by the expression
$$f(p)={p\over\pi}\zeta_{1}\left({{i\pi}\over{2a}}\right)-\zeta_{1}
\left({{ip}\over{2a}}\right),\eqno(8)$$
where $\zeta_{1}$ is the Weierstrass zeta function with quasiperiods 1 and
$\omega=i\pi/a$. It admits also the representation [6]
$$f(p)=ia\cot{p\over 2}-a\sum_{n=1}^{\infty}\left[
\coth\left({{ip}\over 2}+an\right)+
\coth\left({{ip}\over 2}-an\right)\right].\eqno(9)$$
From these representations, one can easily find that $f$ is odd and double
quasiperiodic in $p$,
$$f(p)=-f(-p),\eqno(10)$$
$$f(p+2\pi)=f(p),\eqno(11)$$
$$f(p+2ia)=f(p)+2a.\eqno(12)$$
The derivative $f'(p)$ is double periodic and  due to (11),
(12) determined by its restriction to the fundamental domain
$$D:\quad 0\leq \Re e\, p<2\pi,\quad -a<\Im m\, p\leq a.$$
The representations of $f'(p)$ are
$$f'(p)={1\over\pi}\zeta_{1}\left({{i\pi}\over{2a}}\right)+{i\over{2a}}
\wp_{1}\left({{ip}\over{2a}}\right)$$
$$=-{{ia}\over{2\sin^2{p\over 2}}}
+{{ia}\over 2}\sum_{n\neq 0}{1\over{\sinh^2 \left({{ip}\over 2}
+an\right)}},\eqno(13)$$
where $\wp_{1}$ is the Weierstrass $\wp$-function with periods $(1,\omega)$.

If $p_{1},p_{2}$ are real, the wave function (5) describes scattering of 
magnons. In the bound state, the wave function must vanish as
$\vert n_{1}-n_{2}\vert\to\infty$. It means that $p_{1}$ and $p_{2}$ should be
complex with $P=p_{1}+p_{2}$ real. The simplest possibility is given by the
choice
$$p_{1}={P\over 2}+iq,\quad p_{2}={P\over 2}-iq,\eqno(14)$$
where $q$ is real. One can always choose $q>0$ for convenience. Then vanishing
of $\psi(n_{1},n_{2})$ as $\vert n_{1}-n_{2}\vert\to\infty$ is equivalent to
the condition
$$\coth\gamma(p_{1},p_{2})={{f(p_{1})-f(p_{2})}\over{2a}}=1.\eqno(15)$$
Moreover, one can choose $q\leq a$. At first sight, the interval $a<q\leq 2a$
should be also taken into consideration but one can see that in this case
the rearrangement $p_{1}\to p_{2}+2ia$, $p_{2}\to p_{1}-2ia$, $q\to 2a-q$
 returns the
problem to the interval $0\leq q< a$ due to quasiperiodicity property (12).

The equation (15) under the condition (14) can be rewritten in more detailed
form,
$$F_{P}(q)=1-{1\over 2a}
\left[{{2iq}\over\pi}\zeta_{1}\left({{i\pi}\over{2a}}\right)-
\zeta_{1}\left({{iP}\over{4a}}-{q\over{2a}}\right)+
\zeta_{1}\left({{iP}\over{4a}}+{q\over{2a}}\right)\right]=0.\eqno(16)$$
At fixed real $P$ and real $q$, the function $F_{P}(q)$ is real. In the limit
$a\to\infty$ it reads
$$\tilde F_{P}(q)=1-{i\over 2}\left[
\cot\left({P\over 4}+{{iq}\over 2}\right)-
\cot\left({P\over 4}-{{iq}\over 2}\right)\right]=
{{e^{-q}-\cos P/2}\over{\cosh q-\cos P/2}}.$$
One can see that for each $-\pi\leq P\leq \pi$ there is just one bound state
 with the
solution $q=-\ln\cos{P\over 2}$ of the equation $\tilde F_{P}(q)=0$.
 In the case of finite $a$, which is
of our interest, the situation becomes more involved. One has to investigate 
zeroes of transcendental function $F_{P}(q)$ (16). To this end, note that
(10) and (12) imply the following properties of $F_{P}(q)$,
$$F_{P}(0)=1,\eqno(17)$$
$$F_{P}(q)=-F_{P}(2a-q),\quad F_{P}(q)=-F_{P}(-q)+2.\eqno(18)$$
This means that $F_{P}(a)=0$, i.e. $q=a$ is a solution to (16)
for all $P$. However, one can see by the direct substitution of (14)
with $q=a$ into (5) that the wave function in this case vanishes identically,
i.e. this zero is unphysical and physical solution must lie in the interval
$0<q<a$. Let us find the existence criterium for it.

Note first that due to (17),(18) $F_{P}(0)=1$ and $F_{P}(a)=0$. Then a
nontrivial zero of $F_{P}(q)$ in the interval $0<q<a$ must exist if the
derivative of $F_{P}(q)$ at the point $q=a$ is positive,
$$F'_{P}(a)=-{i\over {\pi a}}\zeta_{1}\left({{i\pi}\over {2a}}\right)
+{1\over {4a^2}}\left[
\wp_{1}\left({{iP}\over{4a}}-{1\over 2}\right)+
\wp_{1}\left({{iP}\over{4a}}+{1\over 2}\right)\right]>0.$$
Since $\wp_{1}(x-1)=\wp_{1}(x)$, this condition can be cast into the form
$$-if'\left({P\over 2}+ia\right)=-{i\over \pi}\zeta_{1}\left({{i\pi}\over
{2a}}\right)+{1\over{2a}}\wp_{1}\left({{iP}\over {4a}}+{1\over 2}\right)$$
$$={a\over 2}\sum_{n=-\infty}^{\infty}{1\over {\sinh^2\left({{iP}\over 4}
+a(n+1/2)\right)}}>0.\eqno(19)$$

Let us prove that the inequality (19) has nontrivial solutions. Note first
that the inequality definitely holds for $P=0$.  On the other hand,
let us estimate the left-hand side at $P=\pi$,
$$\sum_{n=-\infty}^{\infty}{1\over {\sinh^2\left({{i\pi}\over 4}
+a(n+1/2)\right)}}=
\sum_{n=0}^{\infty}\left[
{1\over {\sinh^2\left({{i\pi}\over 4}+a(n+1/2)\right)}}+
{1\over {\sinh^2\left({{i\pi}\over 4}-a(n+1/2)\right)}}\right]$$
$$=-\sum_{n=0}^{\infty}{4\over {\cosh^2 (2n+1)a}}<0.$$
It means that there should be some value $P_{cr}(a)$, $0<P_{cr}(a)<\pi$,
such that inequality (19) is satisfied for all values $0\leq P < P_{cr}$,
since the left-hand side of (19) is a continuous function of $P$.
At these values of the total pseudomomentum there should be at least one
bound state specified by (14),(15). At $P>P_{cr}$, the inequality (19)
does not hold and there are no bound states of the type I since the 
derivative of the left-hand side of (19) with respect to $P$ does not change
sign in the interval $(0,\pi)$ (cf. the known zeroes of function $\wp'_{1}$).

Let us consider another possibility for getting the bound state. To this end,
note that
$$f(p+ia)={p\over \pi}\zeta_{1}\left({{i\pi}\over{2a}}\right)
      +{{ia}\over \pi}\zeta_{1}\left({{i\pi}\over{2a}}\right)
-\zeta_{1}\left({{ip}\over {2a}}-{1\over 2}\right).$$
Taking into account the relations
$$\zeta_{1}\left({{ip}\over {2a}}-{1\over 2}\right)=\zeta_{1}\left(
{{ip}\over {2a}}\right)-\zeta_{1}\left({1\over 2}\right)+
{{\wp'_{1}\left({{ip}\over {2a}}\right)}\over
{2\left(\wp_{1}\left({{ip}\over {2a}}\right)-\wp_{1}\left({1\over 2}\right)
\right)}},$$ 
$${{ia}\over \pi}\zeta_{1}\left({{i\pi}\over {2a}}\right)
+\zeta_{1}\left({1\over 2}\right)=a,$$
one finds that
$$f(p+ia)=a+i\chi(p),\quad f(p-ia)=-a+i\chi(p),\eqno(20)$$
where $\chi(p)$ is real for real $p$. Suppose now that
$$p_{1}=\tilde p_{1}+ia,\quad p_{2}=\tilde p_{2}-ia,\eqno(21)$$
where both $\tilde p_{1}$ and $\tilde p_{2}$ are real, $\tilde p_{1}\neq
\tilde p_{2}$ (the case $\tilde p_{1}=\tilde p_{2}$ corresponds to trivial
solution to the equation (15) mentioned above). Substitution of (21) into (15)
gives with the account of (20)
$$\chi(\tilde p_{1})-\chi(\tilde p_{2})=0.\eqno(22)$$
Note also that
$$\chi(0)=\chi(\pi)=0,$$
$$  \chi'(0)={a\over 2}\sum_{n=-\infty}^{\infty}{1\over {\sinh^2 a\left(n+
{1\over 2}\right)}}>0,\eqno(23)$$
$$\chi'\left({\pi\over 2}\right)
={a\over 2}\sum_{n=-\infty}^{\infty}{1\over {\sinh^2 \left(
{{i\pi}\over 4}+a\left(n+{1\over 2}\right)\right)}}<0.$$
It follows from (23) that there should be some value $\tilde p_{0}$ at
which $\chi(\tilde p)$ has a maximum on the interval $[0,\pi]$ and the
corresponding $\tilde p_{0}'=2\pi-\tilde p_{0}$ at which $\chi(\tilde p)$
has a minimum on the interval $[\pi,2\pi]$. As a matter of fact, $
\tilde p_{0}={{P_{cr}}\over 2}$. Since $f'(p)$ is linearly related to the
Weierstrass function $\wp_{1}$, $\chi'(\tilde p)$ can have only two zeroes
in the fundamental domain, i.e. there are no other extrema of $\chi
(\tilde p)$ on the interval $[0,\pi]$. The presence of a maximum means that
the equation
$$\chi(\tilde p)=\chi_{0}$$
has two distinct real roots if $0\leq \chi_{0}<\chi\left({{P_{cr}}\over 2}
\right)$, ${{P_{cr}}\over 2}<\tilde p_{1}\leq \pi$ and $0\leq \tilde p_{2}
<{{P_{cr}}\over 2}$. These roots serve also as nontrivial solution to the
equation (22) and thus give, via (21), the bound state of type II in which 
the wave function oscillates and decays exponentially as $\vert n_{1}-n_{2}
\vert\to\infty$. For $P_{cr}<P\leq \pi$ such a solution always exists.
Similar solutions corresponding to $-\chi\left({{P_{cr}}\over 2}\right)
< \chi_0 < 0$ can be found with any $-\pi\leq P <-P_{cr}$.
The methods used above do not allow us to say anything definite about the
presence of bound state at $\pi<\vert P\vert \leq 2\pi$
different from the just described ones.

In conclusion, we have proved the existence of two types of bound states
of two magnons in the model with nonlocal spin interaction (1-2). The states
of type I appear for small values of the total pseudomomentum $P$,
$0<\vert P\vert <P_{cr}<\pi$. Due to oscillation of terms in the series (7), 
one can state
that the energy for the type I states is lower than the continuum boundary
only for sufficiently small values of $P$. The state of the type II appears
if the total pseudomomentum exceeds the critical value, 
$P_{cr}<\vert P\vert \leq \pi$.
Our treatment has been universal with respect to parameter $a$ in the interval
$0<a<\infty$. In the nearest-neighbor limit $a\to\infty$, the type II states
with complex relative pseudomomentum and oscillating wave function disappear
$(P_{cr}\to\pi$ as it can be seen from (19)) and the result coincides with
the known one. An interesting open problem concerns the presence or absence
of multimagnon bound states. The condition of their existence is given by
the set of transcendental equations of the type (15). At present, we have not
found the way of their analytical investigation. 
\\
\\
{The work is supported by ASCR grant no. 148409 and GACR grant 
no. 202/96/0218.}
\newpage
{\bf References}
\begin{enumerate}
\item
H.A. Bethe. Z.Phys. 71,205 (1931)
\item
C. Majumdar. J.Math.Phys. 10,177 (1969)
\item
I. Ono, S. Mikado and T. Oguchi. J.Phys.Soc.Japan 30,358 (1971)
\item
I.G. Gochev. Teor.Mat.Fiz. 15,120 (1973)
\item
V.I. Inozemtsev. Commun.Math.Phys. 148,359 (1992)
\item
E.T. Whittaker and G.N. Watson. A Course of Modern Analysis.
Cambridge at University Press (1927)
\end{enumerate}
\end{document}